\documentclass[a4]{article}

\usepackage{graphicx}
\usepackage{dcolumn}
\usepackage{amsmath,amsfonts}
\usepackage{bm}
\usepackage{pslatex}
\usepackage[dvips]{epsfig}
\usepackage{amssymb}
\usepackage{ae}

\usepackage{graphicx}

\title{Analytical study of the propagation of acoustic waves in a 1D weakly disordered lattice}
\author{O. Richoux, E. Morand and L. Simon}

\begin{document}

\maketitle

\begin{abstract}
This paper presents an analytical approach of the propagation of an acoustic wave through a normally distributed disordered lattice made up of Helmholtz resonators connected to a cylindrical duct. This approach allows to determine analytically the exact  transmission coefficient of a weakly disordered lattice. Analytical results are compared to a well-known numerical method based on a matrix product. Furthermore, this approach gives an analytical expression of the localization length apart from the Bragg stopband which depends only on the standard deviation of the normal distribution disorder. This expression permits to study on one hand the localization length as a function of both disorder strength and frequency, and on the other hand, the propagation characteristics on the edges of two sorts of stopbands (Bragg and Helmholtz stopbands). Lastly, the value of the localization length inside the Helmholtz stopband is compared to the localization length in the Bragg stopband.
\end{abstract}

\noindent Wave propagation in random media, waveguide, scattering of acoustic waves.

\section{Introduction}

Since 1958 and the work of P. W. Anderson on the localization of wave in a random media \cite{Anderson58}, the propagation of waves in complex media has been at the center of many works. Number of them have been undertaken on the effect of disorder on the wave propagation and many research fields have been concerned. The first studies appeared in solid state physics \cite{Schmidt57,Dyson53} and in the propagation of EM waves in random medium \cite{Tiggelen92,Economou89}. The propagation of classical waves in disordered media \cite{Soukoulis94,Economou90,Sheng90} and the localization of elastic waves \cite{Ottarsson97,Hodges83,Charles06} have been also dealt with theoretical and experimental studies and the applications, for example in acoustics \cite{Sugimoto95,Richoux02b,Richoux07,Sornette92} or in geophysics \cite{Ursin83,Gilbert80} are numerous. In the same time, number of works have proposed experimental results in 1D \cite{Soukoulis94,Richoux02b,Richoux02a,Sugimoto95}, 2D \cite{Ye00,Liu00} and 3D media \cite{Yang02,Page05}. 

Among these numerous references, only a few analytical studies of Anderson localization have been published. First of all, Kholer \textit{et al} studied, in the 70's, the wave propagation in a one-dimensional medium with random index of refraction or a transmission line with random capacitance per unit length using the radiative transport theory \cite{Kholer73}. In this study, the "disorder" is measured by a time-invariant refractive index field. The frequency dependence of the localization length for acoustic and electromagnetic waves in a one-dimensional randomly layered media is also studied analytically \cite{Sheng86b}. An analytical theory of a pulse propagating in a one-dimensional layered media \cite{Burridge88} is also established by using asymptotic methods for stochastic differential equations \cite{Kholer91} and a low frequency limit is considered to study the localization of elastic waves in a plane-stratified media \cite{Kholer96}.

Secondly, the problem of pulse backscattering from a randomly stratified media is considered where the wave speed fluctuations depend on time and on the range coordinate. The disorder is measured by a time-varying refractive index field and the time variation is parametric \cite{Frankenthal03}. Radiation transport equations are used to describe the propagation and the wave localization in time domain \cite{White87}. 

In the same way, some works present analytical results by using approximate methods like the Coherent Potential Approximation \cite{Economou89,Soukoulis94}, in the case of small amount of impurities \cite{Schmidt57,Sebbah93}, for the low frequency case \cite{Chow73} or for asymptotic behavior \cite{Lobkis05}.

Nevertheless, the Transfer Matrix method \cite{Soukoulis94,Richoux02b} is generally used to simulate the wave propagation in a random lattice \cite{Barnes91}.

The present paper proposes a new analytical approach to study the propagation of acoustic waves in a 1D random media. This method is based on a recursive relation describing the wave propagation. It uses the properties of the normal distribution disorder to propose an analytical expression for the transmission coefficient depending on the standard deviation of the disorder distribution and on the number of lattice cells.

The section II presents a general study of the wave propagation in a 1D disordered lattice made up of Helmholtz resonators connected to a main waveguide. The disorder is introduced through the elementary cell length. An exact expression of the transmission coefficient of the lattice is established in the form of a recursive relation. Thanks to the properties of the normal distributed disorder, an analytical expression of the transmission coefficient modulus and of the localization length is proposed. In section III, the results of this analytical model are compared with the results of Monte Carlo simulations based on the Transfer Matrix Method and a discussion is held.

\section{Propagation of acoustic waves in a weakly Gaussian disordered one-dimensional lattice}
\label{sec:Analytical_expression}
\subsection{General study}

\begin{figure}[h!]
\centering
\includegraphics[width=8cm]{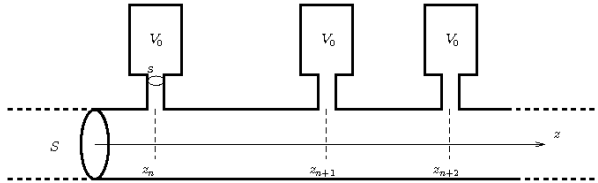}
\caption{\label{figure_resonator} Schematic representation of the Helmholtz resonator lattice made up of a cylindrical waveguide with section $S$ connected to an array of Helmholtz resonator.}
\end{figure}

A one-dimensional lattice is considered made up of an infinitely long cylindrical waveguide (with section $S$) connected to an array of Helmholtz resonators at position $z_n$ (Fig.~\ref{figure_resonator}). The Helmholtz resonators are connected to the cylindrical duct through a pinpoint connection, the radius $r$ of the throat's cross sectional area $s$ of the resonators being assumed to be small compared with the wavelength $\lambda$ of the acoustic wave ($\sqrt{r/\lambda} \ll 1$).

\subsubsection{Model of Helmholtz resonator}

\begin{figure}[h!]
\centering
\includegraphics[width=6cm]{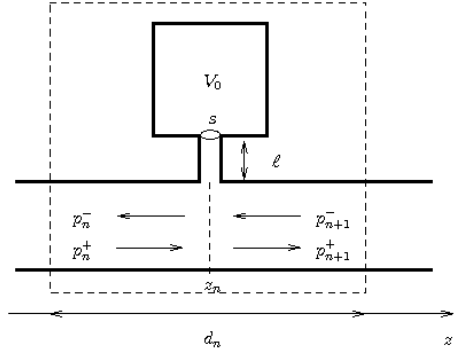}
\caption{\label{fig:resonator} Elementary cell $n$ made up of an Helmholtz resonator and a part of the waveguide.}
\end{figure}

The Helmholtz resonators are composed of a neck with section $s$ and length $\ell$ connected to a volume $V_0$ (Fig.~\ref{fig:resonator}). A simple model of the Helmholtz resonator requires the following assumptions: i) the pressure inside the volume $V_0$ is spatially uniform, ii) the fluid in the neck moves like a solid piston. In this case, the air enclosed in the resonator acts as a spring for the lumped mass of air moving within the neck. We can furthermore consider the loss in the volume $V_0$ by means of a dashpot. In these conditions, the relative change of the pressure $p(t)=p e^{j \omega t}$ in the volume $V_0$ due to a small displacement $x(t)$ of the air in the neck induces a restoring force $F(t)$
\begin{equation}
F(t)= p(t) s = -\frac{\rho c^2 s^2}{V_0}x(t),
\label{force_resonator}
\end{equation}
where $\rho$ is the air density and $c$ the sound velocity \cite{Richoux07}. The spring-dashpot force is considered here as linear. For a monochromatic wave (with angular frequency $\omega$), the displacement $x(t)=x e^{j \omega t}$ of the air  is related to the acoustic velocity in the neck $v_H(t)=v e^{j \omega t}$ by the relation $v_H(t)=j \omega x(t)$ and the Euler relation applied to the air mass $m=\rho \ell_e s$ (where $\ell_e$ is the effective neck length) submitted to the harmonic force $p/(\rho \ell_e) e^{j\omega t}$ gives
\begin{equation}
j \omega v + \alpha v + \frac{\omega_0^2}{j \omega} v = p/(\rho \ell_e),
\label{eq_resonator}
\end{equation}
where $\omega_0=c\sqrt{\frac{s}{V_0\ell_e}}$ is the eigenfrequency of the Helmholtz resonator and $\alpha$ represents the loss in the resonator.

\subsubsection{Equation propagation in the Helmholtz resonators lattice}

For a monochromatic acoustic wave with a frequency below the cut-off frequency of the waveguide, the acoustic pressure and velocity along the waveguide are respectively written $p(z,t)=p(z) e^{j\omega t}$ and $v(z,t)=v(z) e^{j\omega t}$. The amplitude $p(z)$ and $v(z)$ are related by an impedance relation. At each connection between the waveguide and a resonator, the wave impedance and the acoustic velocity are discontinuous. For $z_{n-1}<z<z_n$ (Fig.~\ref{figure_resonator}), the pressure and the acoustic velocity are denoted by $p_n(z)$ and $v_n(z)$ and for $z_n<z<z_{n+1}$, the pressure and the acoustic velocity are denoted by $p_{n+1}(z)$ and $v_{n+1}(z)$. The acoustic pressure is written as a linear conbination of forward and backward waves (Fig.~\ref{fig:resonator}) as
\begin{equation}
p_n(z) = p_n^+ e^{jk(z-z_n)} + p_n^- e^{-jk(z-z_n)},
\label{pressure}
\end{equation}
where $k$ is the wave number defined as $k=\omega/c$.
The continuity of the acoustic pressure at $z=z_n$ yields
\begin{equation}
p_n^+ +p_n^-=p_{n+1}^+ +p_{n+1}^-=p.
\label{continuity_pressure}
\end{equation}
Moreover, the conservation of the mass flux at $z=z_n$ leads to
\begin{equation}
s v = \frac{S}{\rho c}(p_{n+1}^+ - p_{n+1}^- + p_{n}^- - p_{n}^+).
\label{continuity_flow}
\end{equation}
Using the Eqs.~(\ref{eq_resonator}), (\ref{continuity_pressure}) and (\ref{continuity_flow}), a relation between the amplitude of waves across the junction $z_n$ can be written with a matrix formalism :
\begin{equation}
\left(
\begin{array}{c}
p_{n+1}^{+}    \\
p_{n+1}^{-}    \\
\end{array}
\right)
\left(
\begin{array}{cc}
1-w  &  -w   \\
w    &  1+w  \\  
\end{array}
\right)
=
\left(
\begin{array}{c}
p_n^{+}    \\
p_n^{-}    \\
\end{array}
\right),
\end{equation}
where $\beta=\frac{cs}{2 l_e S}$ and $w=\frac{\beta j \omega}{(\omega_0^{2}-\omega^{2})+\alpha j \omega}$. 

We consider here  the n$^{th}$ cell (Fig.~\ref{fig:resonator}) made up of a resonator connected to the middle of a pipe of length $d_n$ and we set $P_{n+1}^{+}$, $P_{n}^{+}$, $P_{n+1}^{-}$ and $P_{n}^{-}$ the amplitudes of forward and backward waves at the two open ends of the n$^{th}$ cell. Then, the relation between the pressure at $z=z_n+d_n/2$ and the pressure at $z=z_n-d_n/2$ can be expressed as 
\begin{equation}
\left(
\begin{array}{c}
P_{n+1}^{+}    \\
P_{n+1}^{-}    \\
\end{array}
\right)
=\left(
\begin{array}{cc}
h_n(1-w)    &  -w      \\
w         &  \frac{1}{h_n}(1+w) \\  
\end{array}
\right)
\left(
\begin{array}{c}
P_{n}^{+}    \\
P_{n}^{-}    \\
\end{array}
\right)=
M_n
\left(
\begin{array}{c}
P_{n}^{+}    \\
P_{n}^{-}    \\
\end{array}
\right),
\end{equation}
where $h_n = e^{-j k d_n}$.

The 1D lattice is made up of $N$ elementary cells embedded in an infinite waveguide. The amplitude of the incident wave on the lattice is noted $P_i$, the amplitude of the reflected wave, $P_r$, and the amplitude of the transmitted wave $P_t$. Using the symmetry of the lattice, the relation describing the propagation of a monochromatic acoustic wave through the lattice is consequently written as 
\begin{equation}
\label{eq:transmission_lattice}
\left(
\begin{array}{c}
P_{r} \\
P_{i} \\
\end{array}
\right)
=M_1 M_2 \dots M_n \dots M_N
\left(
\begin{array}{c}
0     \\
P_{t} \\
\end{array}
\right).
\end{equation}
According to the Furstenberg's theorem \cite{Furstenberg63} concerning the product of random matrix, Eq.~(\ref{eq:transmission_lattice}) describes the propagation in a random media (each matrix $M_n$ is different) showing localization phenomenon \cite{Kirkpatrick85}.

\subsection{Recursive relation for the transmission coefficient}

The reflexion coefficient of the lattice $R_N$ and transmission coefficient $T_N$ being defined by 
\begin{equation}
R_N=\frac{P_r}{P_i} \mbox{ and } T_N=\frac{P_t}{P_i},
\end{equation}
the relation (\ref{eq:transmission_lattice}) is now written
\begin{equation}
\label{eq:coef_trans_reflex}
\left(
\begin{array}{c}
\frac{R_N}{T_N} \\
\frac{1}{T_N} \\
\end{array}
\right)
=M_1 M_2 \dots M_N
\left(
\begin{array}{c}
0 \\
1 \\
\end{array}
\right).
\end{equation}
Then, if we define the matrix $\mathcal{M}$ as
\begin{equation}
\mathcal{M}=M_1 M_2 \dots M_{N-2} =
\left(
\begin{array}{cc}
m_1  &  m_2   \\
m_3  &  m_4   \\  
\end{array}
\right),
\end{equation}
the relation 
\begin{equation}
\left(
\begin{array}{c}
\frac{R_{N-2}}{T_{N-2}} \\
\frac{1}{T_{N-2}} \\
\end{array}
\right)
=\mathcal{M}
\left(
\begin{array}{c}
0 \\
1 \\
\end{array}
\right)
\end{equation}
and the Eq.~(\ref{eq:coef_trans_reflex}) leads to $m_4 = \frac{1}{T_{N-2}}$. As a consequence, the relation
\begin{equation}
\left(
\begin{array}{c}
\frac{R_{N-1}}{T_{N-1}} \\
\frac{1}{T_{N-1}} \\
\end{array}
\right)
=\mathcal{M} M_{N-1}
\left(
\begin{array}{c}
0 \\
1 \\
\end{array}
\right),
\end{equation}
allows us to express $m_3$ as 
\begin{equation}
m_3 = w \left[ \frac{1}{h_{N-1}}(1+w)\frac{1}{T_{N-2}} - \frac{1}{T_{N-1}} \right]. 
\end{equation}

Finally, using 
\begin{equation}
\left(
\begin{array}{c}
\frac{R_N}{T_N} \\
\frac{1}{T_N} \\
\end{array}
\right)
=\mathcal{M} M_{N-1} M_N
\left(
\begin{array}{c}
0 \\
1 \\
\end{array}
\right),
\end{equation}
a recursive relation between $T_N$, $T_{N-1}$ and $T_{N-2}$ can be written
\begin{equation}
\label{recursive_Trans}
\frac{1}{T_N} = 
\left[ h_{N-1}(1-w)+\frac{1}{h_N}(1+w) \right] \frac{1}{T_{N-1}} -\frac{1}{T_{N-2}}.
\end{equation}
Thus, the transmission coefficient $T_N$ can be calculated from $T_0=1$ and $\displaystyle T_1=\frac{h_1}{1+w}$. In the following, results of a numerical simulation of Eq.~(\ref{recursive_Trans}) are considered as "reference results".

\subsection{Analytical expression for the transmission coefficient}

We now consider a weakly normally distributed disordered lattice with the following probability density function of the cell length $d_n$
\begin{equation}
P_{d_n}(x)=\frac{1}{\sigma \sqrt{2\pi}}e^{-\frac{(x-d)^2}{2\sigma ^2}},
\end{equation}
where $d$ is the mean value of the cell length and $\sigma^2$ is the variance. Regarding the values of $d$ and $\sigma$ used in the simulation in section \ref{sec:results}, we consider here that $P_{d_n}(x \in ]-\infty;0])\simeq0$. If we set 
\begin{equation}
u_N=\frac{h_1 h_2 \dots h_N}{T_N},
\end{equation}
the discrete recursive relation (\ref{recursive_Trans}) can then be rewritten as
\begin{equation}
u_N=[h_{N-1}h_N(1-w)+(1+w)] u_{N-1}-[h_{N-1}h_N]u_{N-2},
\label{eq2}
\end{equation}
with $u_0=1$ and $u_1=1+w$. 

The value of $u_N$ can then be calculated using Eq.~(\ref{eq2}). Defining $h=e^{-jk d}$ and using 
$\langle h_N \rangle =h e^{-\frac{\sigma^2 k^2}{2}}$ and $\langle h_N^2 \rangle~=~h^2e^{-2 \sigma^2 k^2}$, the mean value of $u_N$ and $h_N u_N$ are
\begin{equation}
\label{moy_uN}
\langle u_N \rangle  =  h e^{-\frac{\sigma^2 k^2}{2}}(1-w)\langle h_{N-1} u_{N-1}\rangle+(1+w) \langle u_{N-1}\rangle
-h^2 e^{-\sigma^2 k^2} \langle u_{N-2}\rangle,
\end{equation}
and
\begin{equation}
\label{moy_hN_uN}
\langle h_N u_N\rangle =  h^2 e^{-2\sigma^2 k^2}(1-w)\langle h_{N-1}u_{N-1} \rangle
+ he^{-\frac{\sigma^2 k^2}{2}}(1+w) \langle u_{N-1}\rangle -h^3e^{-\frac{5\sigma^2 k^2}{2}} \langle u_{N-2}\rangle.
\end{equation}
Finally, rewritting Eq.~(\ref{moy_hN_uN}) as follows,
\begin{equation}
\langle h_N u_N\rangle  = he^{-\frac{3\sigma^2 k^2}{2}}\langle u_N\rangle 
-he^{-\frac{\sigma^2 k^2}{2}}(1+w)(e^{-\sigma^2 k^2}-1) \langle u_{N-1}\rangle,
\end{equation}
leads to a recursive relation concerning the mean value of $u_N$ as
\begin{equation}
\langle u_N \rangle  =  
\left[ h^2e^{-2\sigma^2 k^2}(1-w) +(1+w) \right] \langle u_{N-1}\rangle
-  h^2e^{-2\sigma^2 k^2} \left[ 1+(e^{\sigma^2 k^2}-1)w^2 \right] \langle u_{N-2}\rangle.
\end{equation}
The solution of this second order recurrence equation can be obtained from the roots $r_1$ and $r_2$ of the associated quadratic equation by the form below
\begin{equation}
\langle u_N \rangle = C_1 r_1^N +C_2 r_2^N,
\end{equation}
where
\begin{eqnarray}
r_1 & = & h \left( A + \sqrt{A^2-B} \right), \\
r_2 & = & h \left( A - \sqrt{A^2-B} \right), \\
C_1 & = & \frac{(1+w)-r_2}{r_1 -r_2}, \\
C_2 & = & \frac{(1+w)-r_1}{r_2 -r_1},
\end{eqnarray}
with 
\begin{eqnarray}
A & = & \frac{1}{2} \left[ h(1-w)e^{-2\sigma^2 k^2} +\frac{1}{h}(1+w) \right] \label{eq:A}, \\
B & = & e^{-2\sigma^2 k^2} \left[ 1+(e^{\sigma^2 k^2}-1)w^2 \right] \label{eq:B}. 
\end{eqnarray}

\noindent Finally, the value of $|T_N|$ can be calculated from the expected value of $u_N$ (in the case of a weak Gaussian disorder)
\begin{equation}
\langle |T_N| \rangle \simeq  \frac{1}{\vert\langle u_N \rangle \vert} = \frac{1}{\vert C_1 r_1^N +C_2 r_2^N \vert}.
\label{eq3}
\end{equation}
The Eq.~(\ref{eq3}) is, in a weak disorder approximation, an analytical solution for the transmission coefficient modulus of a disordered 1D lattice depending directly on the "strength" of the disorder $\sigma$ and on the number of cells $N$. 

\subsection{Analytical expression for the localization length}

\noindent In this section, the acoustic attenuation in the Helmholtz resonator is neglected, so that $\alpha=0$ and $w=\frac{\beta j \omega}{\omega_{o}^{2}-\omega^{2}}$. We furthermore suppose that $\omega_0 \leq \frac{\pi c}{d}$ which means that the Helmholtz resonance frequency is smaller than the Bragg frequency $f_B=\pi c/d$.\\

\noindent Defining $\eta=\frac{\beta \omega}{\omega_{o}^{2}-\omega^{2}}$ and $\delta=\frac{1}{2}(e^{-2 \sigma^2 k^2}-1)$, Eqs.~(\ref{eq:A}) and (\ref{eq:B}) become
\begin{eqnarray}
A&=&\Gamma_1+(\Gamma_1-j\Gamma_2)\delta,\label{eq:Abis}\\
B&=&1+(2+\eta^2)\delta,\label{eq:Bbis}
\end{eqnarray}
where 
\begin{equation}
\Gamma_1 = \cos(kd)-\eta\sin(kd)
\end{equation}
and
\begin{equation}
\Gamma_2  =  \sin(kd)+\eta\cos(kd).
\end{equation}

\noindent A first order expansion in $\delta$ gives for $\sqrt{A^2-B}$,
\begin{equation}
\label{eq:A2-B}
\sqrt{A^2-B} \simeq \sqrt{(\Gamma_1^2-1)+(2\Gamma_1^2-2-\eta^2-2j\Gamma_1 \Gamma_2)\delta}.
\end{equation}
Outside the stopbands of the ordered case, $\Gamma_1^2-1<0$, and Eq.~(\ref{eq:A2-B}) may be approximated by
\begin{equation}
\sqrt{A^2-B}\simeq \left(j\sqrt{1-\Gamma_1^2}\right)
+\left(\frac{-\Gamma_1 \Gamma_2+j(1-\Gamma_1^2+\frac{\eta^2}{2})}{\sqrt{1-\Gamma_1^2}}\right)\delta.
\end{equation}

\noindent Lastly, this expression leads to the modulus of the roots $r_1$ and $r_2$ as
\begin{equation}
|r_{1,2}|\simeq 1+\left(1+\frac{\eta^2}{2}\mp \frac{\Gamma_2}{\sqrt{1-\Gamma_1^2}} \right)\delta.
\end{equation}
Because the modulus of one of the roots is less than $1$, it can be neglected and, finally, the expression for the value of the transmission coefficient outside the stopbands of the ordered case is
\begin{equation}
\langle |T_N| \rangle \simeq \exp \left[ -\delta N \left( 1+\frac{\eta^2}{2}
-\sqrt{1+\frac{\eta^2}{1-\Gamma_1^2}}\right) \right].
\label{eq5}
\end{equation}
A comparison between Eq.~(\ref{eq5}) and the relation
\begin{equation}
\label{eq:T_loc}
\langle \vert T \vert \rangle =e^{-L/\xi},
\end{equation}
giving the transmission coefficient of a disrodered lattice as a function of the localization length $\xi$ and the lattice length $L$ \cite{Knapp89}, leads to the asymptotical expression of the adimensionnal localization length :
\begin{equation}
\label{eq:localization_length}
\xi/d = \lim_{N \to + \infty}\frac{-N}{\ln \vert T_N \vert}=\frac{1}{\delta} \left( 1+\frac{\eta^2}{2}
-\sqrt{1+\frac{\eta^2}{1-\Gamma_1^2}}\right)^{-1}.
\end{equation}
On one hand, the localization length is independent of the lattice length. On the other hand, conversely to the use of a Monte Carlo simulation, the calculation of the localization length $\xi$, may be easily achieved with a low computational cost thank to Eq.~(\ref{eq:localization_length}).

\section{Results and discussion}
\label{sec:results}

\noindent By using the recursive relation (\ref{recursive_Trans}), the propagation of an acoustic wave through a lattice made up of $N$ Helmholtz resonators connected to an infinite cylindrical waveguide with surafe area $S$ is simulated. Each cell of the lattice has a mean length $d$. An Helmholtz resonator consists of a cylindrical volume $V_0$ associated to a neck made up of a cylindrical tube of length $\ell$ and of surface area $s$.

\subsection{Transmission coefficient of a weakly Gaussian disordered 1D lattice}

\begin{figure}[h!]
\centering
\includegraphics[width=8cm]{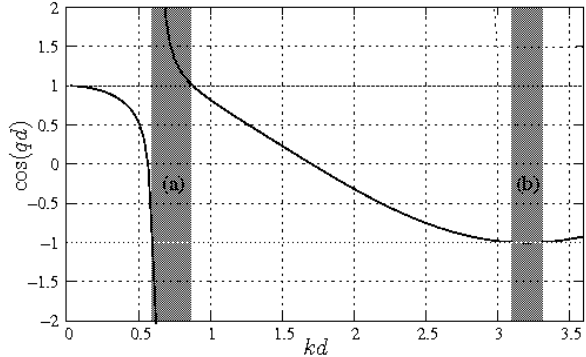}
\caption{\label{fig:ordonne} Dispersion relation of the ordered lattice (Eq. \ref{eq:dispersion}). The grey region show the stopbands of the lattice : resonance stopband \textbf{(a)} and Bragg stopband \textbf{(b)}.}
\end{figure}

The Fig.~(\ref{fig:ordonne}) shows the well-known dispersion relation of the ordered lattice described above, where all the cells have the same length $d$. The relation dispersion is given by \cite{Richoux07}
\begin{equation}
\label{eq:dispersion}
\cos (qd) = \cos (kd) - \frac{1}{2} S_q D_q \frac{1}{kd}\frac{1}{1-k_0^2/k^2} \sin(kd),
\end{equation}
where $q$ is called the Bloch wave number, $S_q=s/S$, $D_q=d/\ell_e$ and $k_0$ is defined by $k_0=\omega_O/c$. In the following, the lattice characteritics are chosen as $S_q~=~7.84~\times~10^{-2}$, $D_q~=~5$ and $k_0~d~=~0.62$.

The dispersion relation exhibits the peculiar characteristic of filters marked by forbidden frequencies or gaps or stopbands (marked in grey on Fig.~\ref{fig:ordonne}) and passbands in the frequency domain which result from the resonances and the periodic arrangements of the medium. When the relation $\vert \cos(qd) \vert \leq 1$ is satisfied, the waves are within a passband and travel freely in the duct. On the contrary, when  $\vert \cos(qd) \vert > 1$, the waves are in a forbidden band and are spatially damped (i.e. evanescent waves). In the lattice described above, two kinds of stopband appear in the band structure : one is due to the resonance of the scatterers (Helmholtz resonators), called resonance stopband or Helmholtz stopband (marked by \textbf{(a)} on Fig.~\ref{fig:ordonne}) and the other is due to the periodicity of the lattice called Bragg stopband (marked by \textbf{(b)} on Fig.~\ref{fig:ordonne}).

\begin{figure}[h!]
\centering
\includegraphics[width=8cm]{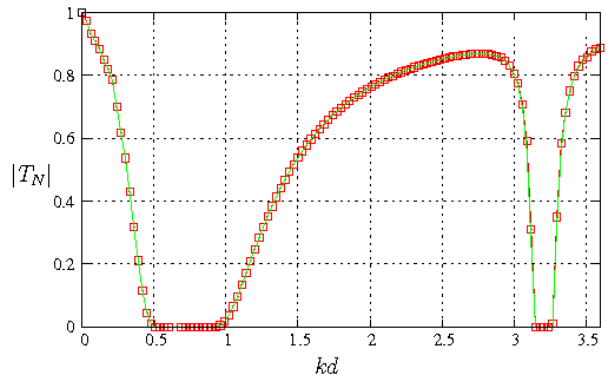}
\caption{\label{fig:tau50} Modulus of the transmission coefficient of a weakly Gaussian disordered lattice for $\sigma = d/50$ : -- analytical calculus (from Eq.~(\ref{eq3})), $\square$ simulation (from Eq.~(\ref{recursive_Trans})).}
\end{figure}

\begin{figure}[h!]
\centering
\includegraphics[width=8cm]{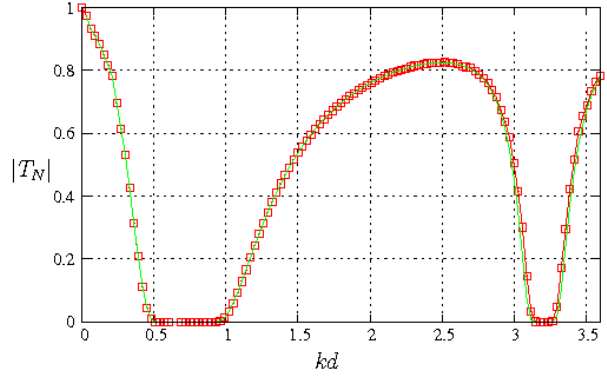}
\caption{\label{fig:tau20} Modulus of the transmission coefficient of a weakly Gaussian disordered lattice for $\sigma = d/20$ : -- analytical calculus (from Eq.~(\ref{eq3})), $\square$ simulation (from Eq.~(\ref{recursive_Trans})).}
\end{figure}

\noindent In Figs.~\ref{fig:tau50} and \ref{fig:tau20}, the moduli of the transmission coefficient of a disordered lattice, determined by a Monte Carlo simulation (based on the Eq.~(\ref{recursive_Trans})) and by the analytical model (from Eq.~(\ref{eq3})), are compared. The Gaussian random sequence used to built the disordered lattice is characterized by a standard deviation $\sigma = d/50$ for the Fig.~\ref{fig:tau50} and $\sigma = d/20$ for the Fig.~\ref{fig:tau20}, where $d$ is the mean cell length. The Monte Carlo simulation of the transmission coefficient is the result of $1000$ realizations and is computed for a lattice made up of $200$ cells. Assuming that $d_n$ form an ergodic sequence, the transmission coefficient is also ergodic \cite{Ottarsson97}. As a consequence, estimating the transmission coefficient of a lattice made up of $N$ elementary cells with $1000$ realizations is asymptotically equivalent to estimating the transmission coefficient of a lattice made up of $1000 \,N$ elementary cells. The comparison of these two results shows a very good agreement whatever the disorder intensity. 

\noindent Firstly, the location in the frequency domain of the stopbands determined with the numerical and analytical methods is in good agreement with the ordered case result. The widths of the stopband in the disordered cases (defined for $|T|$ close to zero) are increasing with the disorder intensity and, as expected, the disorder on the cell length clearly acts on the Bragg stopbands width \cite{Richoux06}. Around the Helmholtz stopband (for $0.3 < kd < 1.1$) and for $\sigma=d/50$ the maximum of the difference between the transmission coefficient modulus estimated by Monte Carlo simulation and calculated with the analytical model is $3\times10^{-4}$. For $\sigma=d/20$, the maximum of this difference is $1.6\times10^{-3}$.

\begin{figure}[h!]
\centering
\includegraphics[width=6cm]{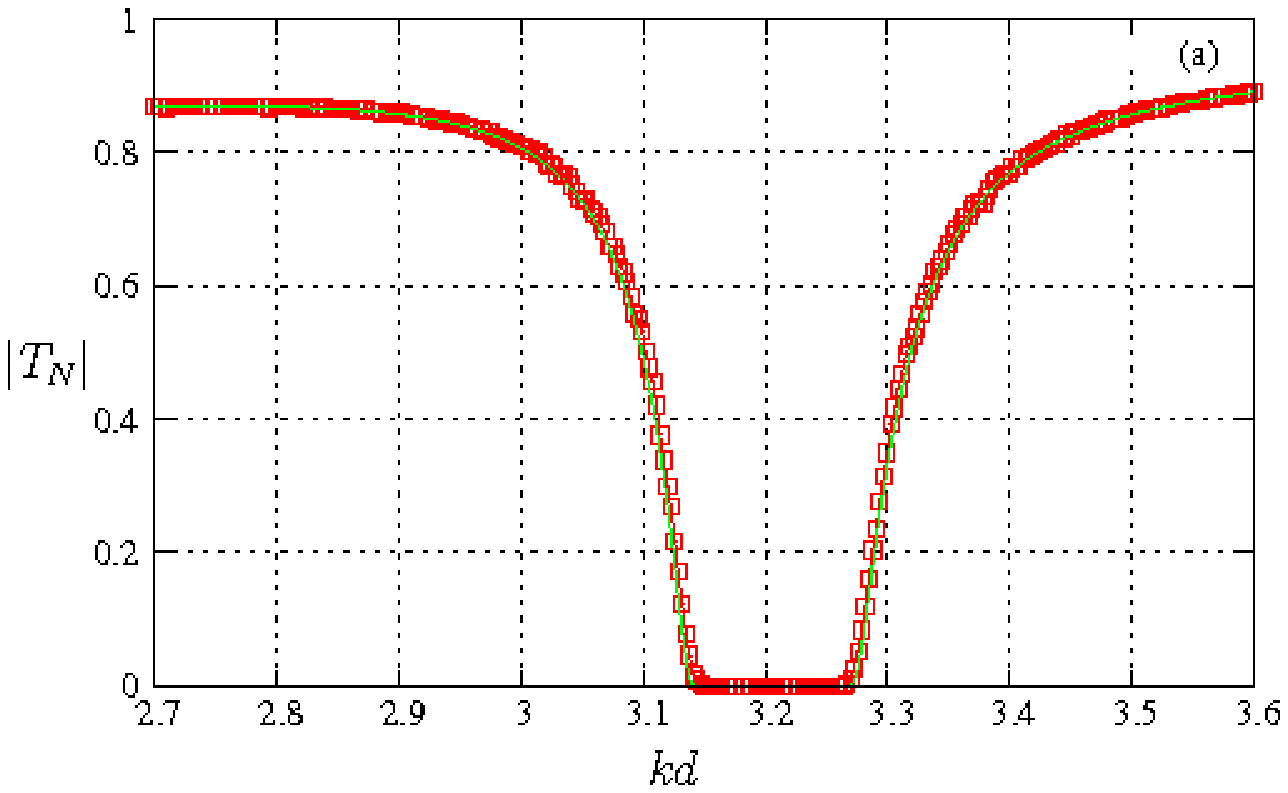} 
\includegraphics[width=6cm]{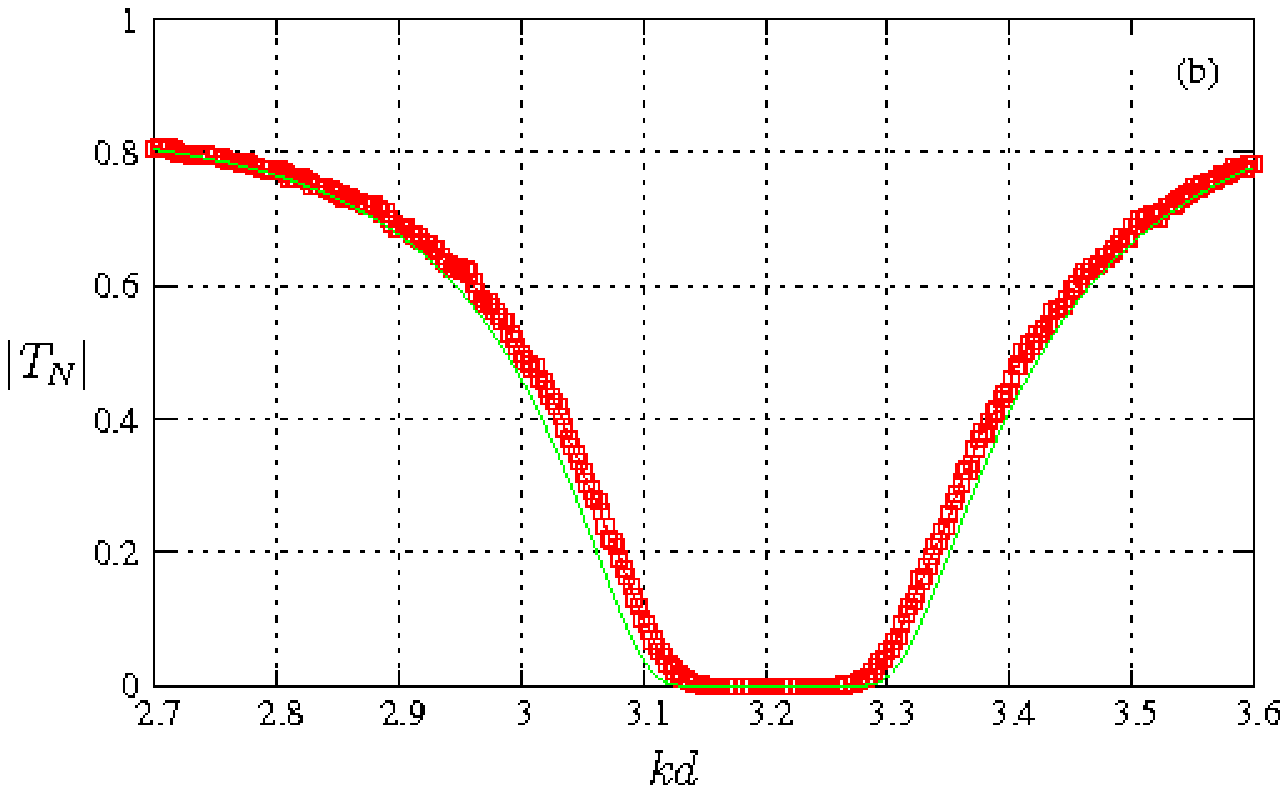}
\caption{\label{fig:zoom_tau} (a) Modulus of the transmission coefficient of a weakly Gaussian disordered lattice for $\sigma = d/50$ and for $2.7< kd < 3.6$ : -- analytical calculus (from Eq.~(\ref{eq3})), $\square$ simulation (from Eq.~(\ref{recursive_Trans})). (b) Modulus of the transmission coefficient of a weakly Gaussian disordered lattice for $\sigma = d/20$ and for $2.7< kd < 3.6$ : -- analytical calculus (from Eq.~(\ref{eq3})), $\square$ simulation (from Eq.~(\ref{recursive_Trans})).}
\end{figure}

Secondly, around the Bragg stopband, the analytical method describes very well the propagation of the wave in the lattice whatever the disorder intensity. Indeed, the comparison between both curves, for Figs.~\ref{fig:tau50} and \ref{fig:tau20}, reveals only a weak difference for the Bragg stopband with $\sigma = d/20$ (Fig.~\ref{fig:tau20}). The Fig.~\ref{fig:zoom_tau} shows a zoom around the Bragg stopband (for $2.7 < kd < 3.6 $) of the transmission coefficient for two different disorder intensities ($\sigma = d/50$ and $\sigma = d/20$). For the low disorder case ($\sigma = d/50$), the difference between the analytical method and the simulation results for the transmission coefficient is always smaller than $0.02$. Nevertheless, with the increase of the disorder strength (for $\sigma = d/20$), the comparison of the two methods shows the validity limit of the analytical model since the maximum of the difference between modulus of analytical transmission coefficient and simulated one reaches $0.05$.  

\noindent According to Eq.~(\ref{eq3}), the modulus of the transmission coefficient of a weakly Gaussian disorder lattice can then be easily estimated from a simple relation depending on the number of the cells, on the physical properties of the lattice and on the standard deviation of the disorder. The computational cost is very low for this new method, since this calculation does not require any matrix product which avoids any divergence problem \cite{Richoux06}.

\subsection{Determination of the localization length}

The analytical model presented in section \ref{sec:Analytical_expression} is used to estimate the localization length of a disordered lattice using Eq.~(\ref{eq:localization_length}) for different disorder intensities and for different frequencies. The lattice under consideration in this study consists of $2000$ cells with a mean length $d$. The relevant adimensional parameter $\nu$ is considered here for measuring the level of disorder, such that
$$
\nu = \frac{6 \sigma}{d}.
$$
As a consequence, a disorder level corresponding to $\nu = 0.5$ leads to about $50$ \% of error on the lattice cell length.\\

\begin{figure}[h!]
\centering
\includegraphics[width=8cm]{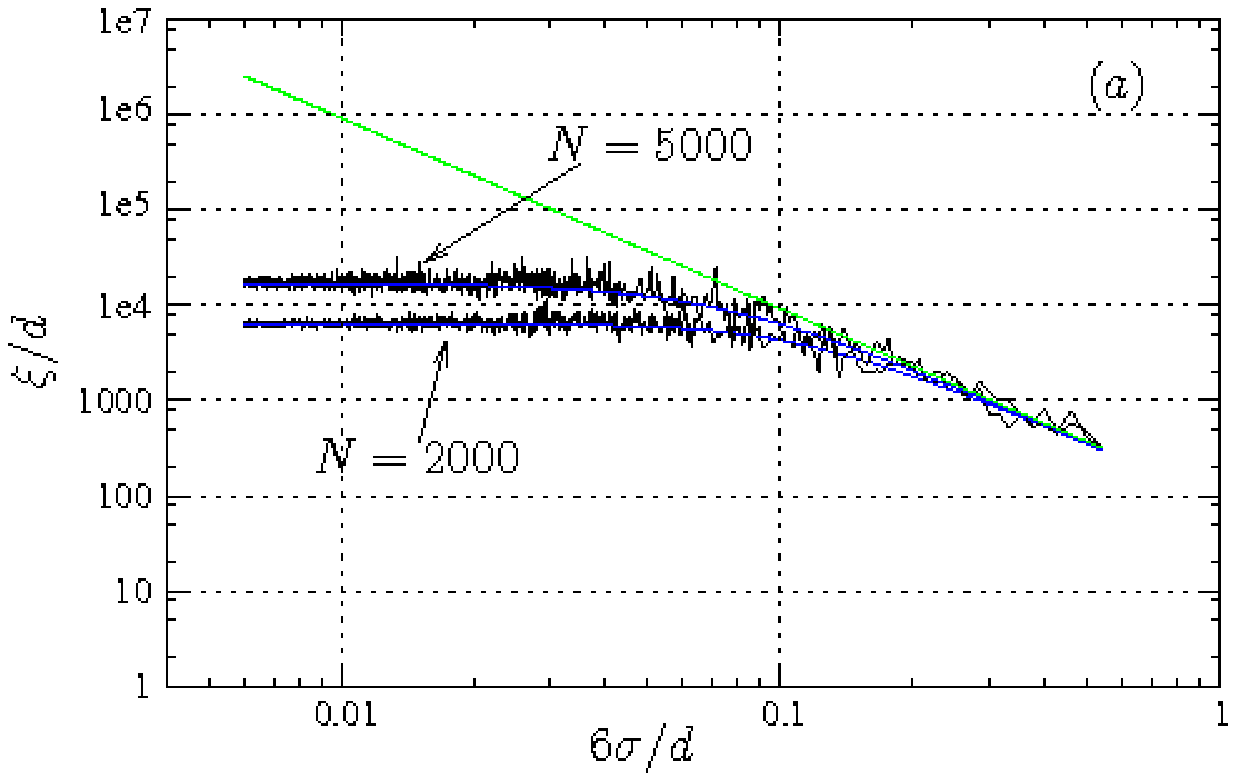}
\includegraphics[width=8cm]{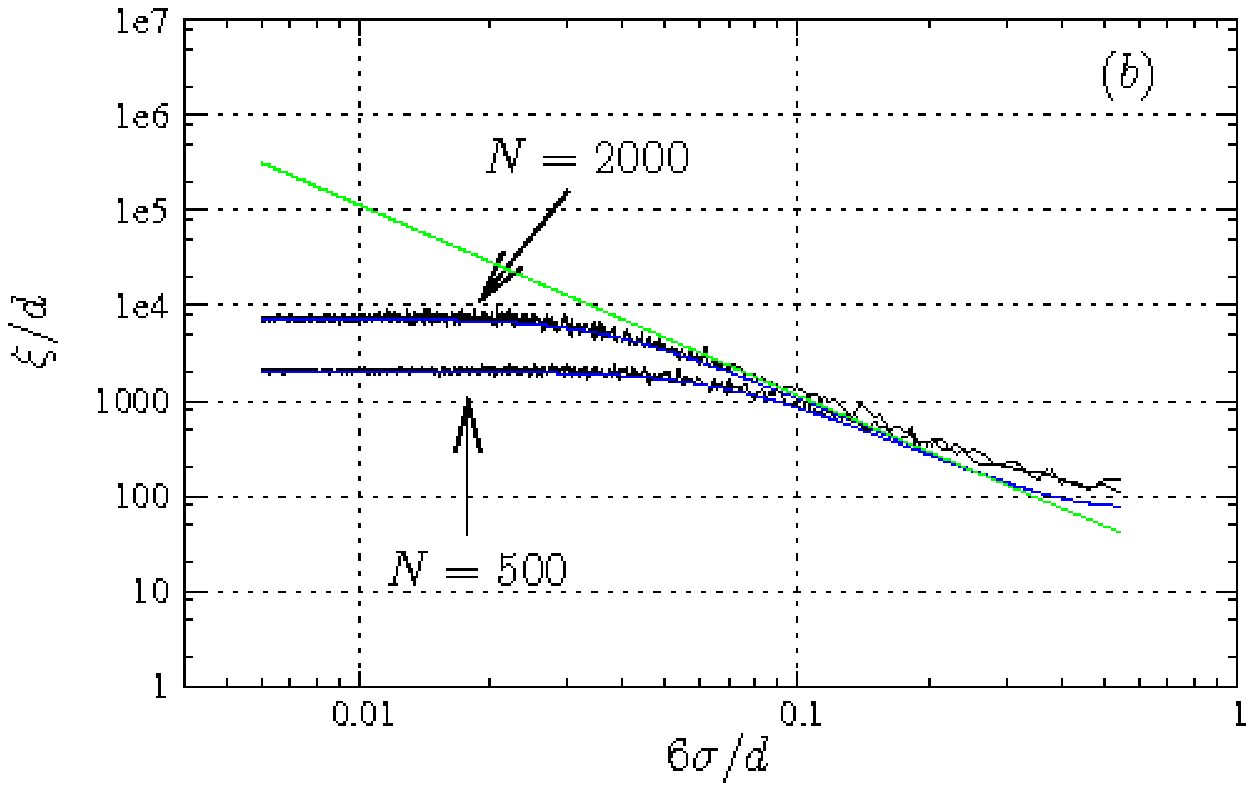}
\caption{\label{fig:long_loc_comp1} (a) Adimensionnal localization length of a Gaussian disordered 1D lattice vs. disorder intensity for a frequency $kd=0.856$. (b) Adimensionnal localization length of a normally distributed disordered 1D lattice vs. disorder intensity for a frequency $kd=3.01$.  The blue line corresponds to the analytical calculus (from the transmission coefficient calculated with the help of Eq.~(\ref{eq3})), the black line to the Monte Carlo simulation results (from the transmission coefficient estimated by Eq.~(\ref{recursive_Trans})) with $50$ realizations for $kd=3.01$ and $10$ realizations for $kd=0.856$ and the green line to the analytical calculus of the localization length (from Eq.~(\ref{eq:localization_length})).}
\end{figure}

\noindent First, the analytical expression of the localization length (Eq.~(\ref{eq:localization_length})) is compared to Monte Carlo simulation results (from Eq.~(\ref{recursive_Trans})) and to the analytical results of Eq.~(\ref{eq3}). For this, the modulus of the transmission coefficient is calculated using Eq.~(\ref{recursive_Trans}) and (\ref{eq3}) and the localization length is estimated using Eq.~(\ref{eq:T_loc}).\\
Figs.~(\ref{fig:long_loc_comp1}a) and~(\ref{fig:long_loc_comp1}b) show this comparison for two different frequencies : (a) $kd=3.01$, i.e., in the low edge of the Bragg stopband, (b) $kd=0.856$, i.e., in the Helmholtz stopband. For each case, two lattices are tested corresponding to $N=500$ and to $N=2000$ cells for $kd=3.01$ and to $N=2000$ and to $N=5000$ cells for $kd=0.856$. For each frequency, the condition $\Gamma_1^2 -1 < 0$ is verified. 

\noindent For $kd=0.856$ the agreement between the different results may be considered as very good. For $N=5000$ and $N=2000$, the Monte Carlo simulation and the calculation of the localization length with the help of the Eq.~(\ref{eq3}) give the same results. The asymptotic limit (for $N \rightarrow +\infty$) of these two methods corresponds to the analytical curve of the localization length (\ref{eq:localization_length}).\\
For $kd=3.01$, the same remark can be made. The agreement between the different results is also very good. The effect of the number of lattice cells on the localization length estimation is clearly shown by comparing the cases $N=500$ and $N=2000$, the case $N=2000$ joining the asymptotic limit (derived from Eq.~(\ref{eq:localization_length})) for a lower value of disorder intensity than the case of $N=500$.

\begin{figure}[h!]
\centering
\includegraphics[width=8cm]{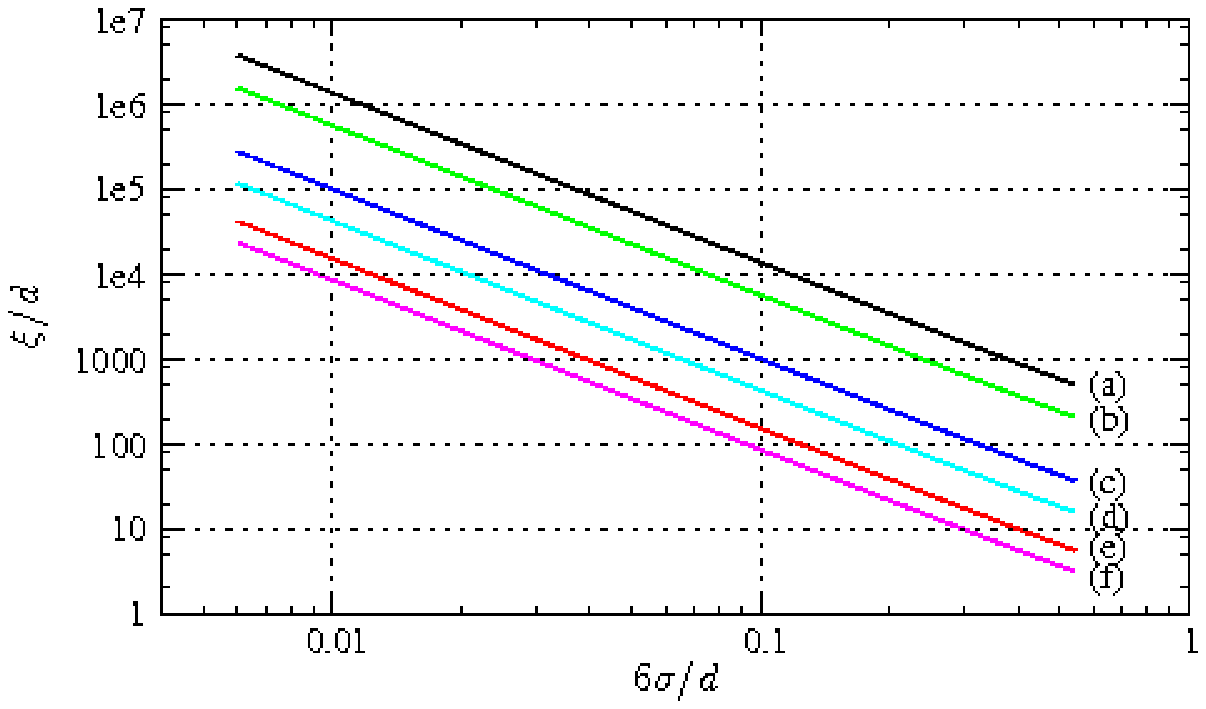}
\caption{\label{fig:long_loc1} Adimensionnal localization length of a Gaussian disordered 1D lattice vs. disorder intensity for an adimensional frequency : (a) $kd=2.83$, (b) $kd=2.92$, (c) $kd=3.01$, (d) $kd=3.03$, (e) $kd=3.041$, (f) $kd=3.043$.}
\end{figure}

Then, Eq. (\ref{eq:localization_length}) can be used to characterize the propagation in a disordered lattice through the localization length. Fig.~\ref{fig:long_loc1} shows the results of such a calculation for $kd=2.83, \mbox{ }2.92, \mbox{ }3.01, \mbox{ }3.03, \mbox{ }3.041, \mbox{ }3.043$. All the frequencies are in the low edge of the first Bragg stopband which is most influenced by the disorder \cite{Richoux06}.

The Fig.~\ref{fig:long_loc1} illustrates, in a $[log;log]$ representation, the adimensionnal localization length as a function of the disorder strength. The influence of the disorder on the localization length is shown. The more disordered the lattice, the smaller the localization length is, whatever the frequency. Regarding the frequencies, the closer they are to the center of the stopband (for $kd$ increasing in the Fig.~\ref{fig:long_loc1}), the smaller the localization length is.

\noindent From the analytical model, it is consequently possible to evaluate the propagation characteritics of the disordered lattice. On one hand, if the lattice length is smaller than the localization length, the wave is localised inside the lattice and the medium is considered as opaque. In the present case, for a disorder level $\nu > 0.3$, all the waves are localized whatever the frequencies. On the contrary, for $\nu \leq 0.025$, all the waves propagate through the lattice.\\

\begin{figure}[h!]
\centering
\includegraphics[width=8cm]{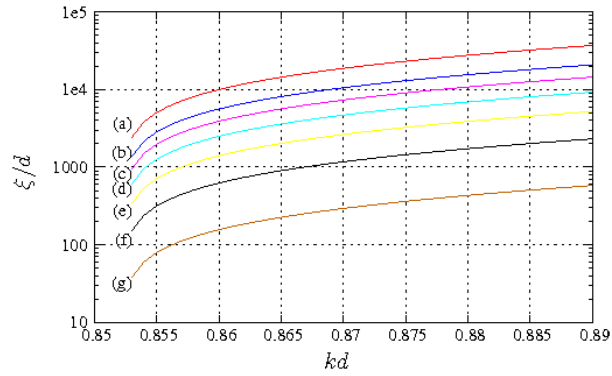}
\caption{\label{fig:long_loc2} Adimensionnal localization length of a Gaussian disordered 1D lattice in function of the adimensional frequency $kd$ for some different values of the disorder level : (a) $6 \sigma/d= 0.05$, (b) $6 \sigma/d= 0.066$, (c) $6 \sigma/d= 0.08$, (d) $6 \sigma/d= 0.1$, (e) $6 \sigma/d= 0.13$, (f) $6 \sigma/d= 0.2$ and (g) $6 \sigma/d= 0.4$. The frequency range is include in the first stopband due to Helmholtz resonance.}
\end{figure}

\noindent Fig.~\ref{fig:long_loc2} presents the adimensionnal localization length as a function of the adimensional frequency $kd$ for different disorder levels in the Helmholtz stopband of the ordered case. Contrary to above, the frequencies are chosen inside the stopband in the disordered lattice case (but outside the stopband of the ordered case where the condition $\Gamma_1^2-1 < 0$ is verified). The localization lengths in this case are comparable with the Bragg stopband case (where the frequency are in the low edge of the stopband) which demonstrates that the Anderson localization can be different, depending on the stopband characteristics. The influence of the disorder level is much more important for frequencies near the Bragg stopband than for frequencies in the Helmholtz stopband. As a consequence, the modulus of the transmission coefficient is not sufficient for studying the propagation wave through a disordered lattice and the analysis of the localization length can bring some precisions about the localization phenomenon. 

\section{Conclusion}

In the present paper, a new method to study analytically the wave propagation in a weakly normally distributed disordered 1D lattice is proposed. The transmission coefficient of the disordered lattice is calculated analytically. The results are in very good agreement with the Monte Carlo simulation results based on a recursive relation applied to the transmission coefficient. Thanks to the analytical formulation, an expression of the localization length independant of the lattice length is also found. The localization length in two kinds of stopband is then characterized and we show that the influence of disorder on the wave propagation is more important near the Bragg stopband than near the Helmholtz stopband.

\end{document}